\PassOptionsToPackage{table,xcdraw}{xcolor}
\documentclass{llncs}
\usepackage[normalem]{ulem}
\usepackage{xcolor}
\usepackage{url}
\usepackage{graphicx}
\usepackage[none]{hyphenat}
\usepackage{calc}
\usepackage{amssymb}
\usepackage{amstext}
\usepackage{amsmath}
\usepackage{stackrel}
\usepackage{multicol}
\usepackage{pslatex}
\usepackage{tabularx}
\usepackage{subfig}
\usepackage{color}
\usepackage{listings}
\usepackage[linesnumbered]{algorithm2e}

\newcommand\Inn{%
  \mathrel{\mathchoice
    {\ooalign{$\displaystyle\subset$\cr\hfil\scalebox{0.8}[1]{$=$}\hfil\cr}}%
    {\ooalign{$\textstyle\subset$\cr\hfil\scalebox{0.8}[1]{$\textstyle=$}\hfil\cr}}%
    {\ooalign{$\scriptstyle\subset$\cr\hfil\scalebox{0.8}[1]{$\scriptstyle=$}\hfil\cr}}%
    {\ooalign{$\scriptscriptstyle\subset$\cr\hfil\scalebox{0.8}[1]{$\scriptscriptstyle=$}\hfil\cr}}%
    }%
}
\newcommand{\HIDE}[1]{}

\newcommand{\comment}[1]{}

\newcommand{\Aplib}{{\sf Aplib}}
\newcommand{\BfAplib}{{\bf Aplib}}
\newcommand{\aplib}{{\sf aplib}}

\begin{document}
\sloppy
\title{An Appraisal Transition System for Event-driven Emotions in Agent-based Player Experience Testing 
\thanks{
This work is funded by EU H2020 Research and Innovation grant 856716, project iv4XR.}
}

\author{
Saba Gholizadeh Ansari \inst{1}\orcidID{ 0000-0002-7135-5605}
\and
   I. S. W. B. Prasetya\inst{1}\orcidID{0000-0002-3421-4635}
   \and Mehdi Dastani\inst{1}\orcidID{0000-0002-4641-4087}
   \and Frank Dignum\inst{2}\orcidID{0000-0002-5103-8127}
   \and Gabriele Keller\inst{1}\orcidID{0000-0003-1442-5387}
}
\institute{
   Utrecht University, Utrecht, the Netherlands,
      \email{s.gholizadehansari@uu.nl}
   \and Umeå University, Umeå, Sweden
}
\maketitle
{\centering \scriptsize \textbf{NOTE:} This is a preprint of the article , accepted in 9th International Workshop on Engineering Multi-Agent Systems (EMAS2021), held as a part of 20th International Conference on Autonomous Agents and Multiagent Systems (AAMAS).}

\begin{abstract}
Player experience (PX) evaluation has become a field of interest in the game industry. Several manual PX techniques have been introduced to assist developers to understand and evaluate the experience of players in computer games. However, automated testing of player experience still needs to be addressed. An automated player experience testing framework would allow designers to evaluate the PX requirements in the early development stages without the necessity of participating human players. In this paper, we propose an automated player experience testing approach by suggesting a formal model of event-based emotions. In particular, we discuss an event-based transition system to formalize relevant emotions using Ortony, Clore, \& Collins (OCC) theory of emotions. A working prototype of the model is integrated on top of {\sf Aplib}, a tactical agent programming library, to create intelligent PX test agents, capable of appraising emotions in a 3D game case study. The results are graphically shown e.g. as heat maps. Emotion visualisation of the test agent would ultimately help game designers in creating content that evokes a certain experience in players. 
\keywords{
  automated player experience testing, emotional modeling of game player, formal model of emotion, intelligent agent, agent-based testing}
\end{abstract}

\section{Introduction}
\label{sec.introduction}
With the growing interest of industry and academia in assessing the quality in-use of a system, product or service, the term \textit{User eXperience} (UX), which refers to quality characteristics related to internal and emotional state of a user, has emerged \cite{peterson2017understanding,rivero2017systematic}. UX evaluations become essential for designers to predict how users would interact with  a system. In the context of computer games, evaluating \textit{player eXperience} (PX) plays an important role to design a well-received game according to  players' preferences and expectations. PX has different dimensions such as flow \cite{procci2012measuring}, immersion \cite{jennett2008measuring} and enjoyment \cite{fang2010development} which need  to be addressed in a game design to evoke certain experience.

To assess the UX quality of a game\HIDE{ in terms of UX}, relatively novel UX evaluation methods such as questionnaire methods, psycho-physiological measurement and eye-tracking 
\HIDE{method} have been used \cite{bernhaupt2015game,rivero2017systematic,vermeeren2010user}. 
Currently, PX testing techniques
not only impose excessive hours of testing but they  might also not be representative enough to cover all player types and their possible emotions towards the game. Despite some attempts towards automation, most of these techniques are either costly or still manually demanding \cite{rivero2017systematic,bernhaupt2015game,vermeeren2010user}. Moreover, similar to UX evaluations in non-game applications, most of PX testing methods measure PX toward the end of the game development \cite{alves2014state,bernhaupt2015game,vermeeren2010user}, so there is still a need for more efficient techniques to do these evaluations in {\em early stages} of game development. This allows PX problems to be addressed early during the development.

All of these factors led us to propose an automated approach for PX testing in computer games; the envisaged main use case is to assist designers in early development phases to develop their games more efficiently.  
To meet this aim, here, we proposes to employ a {\em computational model} of players to automatically assess PX properties of a computer game. Such a model is necessarily tied to cognition and emotion. Additionally, emotions that a player can feel under certain conditions would eventually 
affect their overall experience. We, therefore, suggest to deploy a well-known \textit{theory of emotions} called \textbf{ OCC} \cite{ortony1988cognitive} to facilitate modeling players with respect to their emotions. 

We present a formal model of the appraisal for OCC emotions using an event-based transition system to serve as the foundation of our automated PX testing approach. It deviates from existing formalization e.g. \cite{adam2009logical,gluz2017probabilistic,steunebrink2007logic}; they have never been used in the software engineering (SE) domain.
This might explain why these formal models  have not been utilized for UX/PX testing. 
%
%
A more fundamental reason is that these models are given in the form of BDI\footnote{Belief-Desire-Intention} logic \cite{meyer2015bdi}. Although  expressive, BDI logic is more a reasoning model rather than a computation model. 
In contrast, our formalization is given in terms of a transition system that directly specifies how to compute the emotional state. Having a transition system provides an opportunity for developers to simply deploy the model in their own systems, whereas a BDI-based formal model would also need a BDI reasoning engine before it can be used for computing.
Furthermore, discrete transition systems have been used to do model checking in software for decades. 
This opens a way to express UX/PX properties in e.g. LTL or CTL \cite{baier2008principles} and verify them through model checking or model checking related techniques. 

A prototype implementation of the formal model is also presented in this paper, along with a demonstration of what it can do on a small case study. The prototype of appraisal model is integrated with \Aplib\ \cite{prasetya2020aplib}, a Java library for agent-based game testing, to create an \textbf{emotional test agent} that uses the OCC theory for emotional appraisal to assess PX requirements in games.

The paper is structured as follows: Section \ref{sec.preliminaries} introduces the OCC theory. Section \ref{sec.UXframework} gives an overview of the proposed framework architecture as well as the role of appraisal in PX evaluations. Section \ref{sec.model} details the formal model of appraisal for event-based emotions. Section \ref{sec.UXresult} explains the early results of the framework in a 3D case study.
Section \ref{sec.relatedwork} discusses some related work and finally Section \ref{sec.concl} concludes the paper and presents future work.


\section{OCC theory of emotion}
\label{sec.preliminaries}

\begin{figure}[htbp]
\vspace{-3mm}
\centerline{\includegraphics[width=110mm,height=90mm]{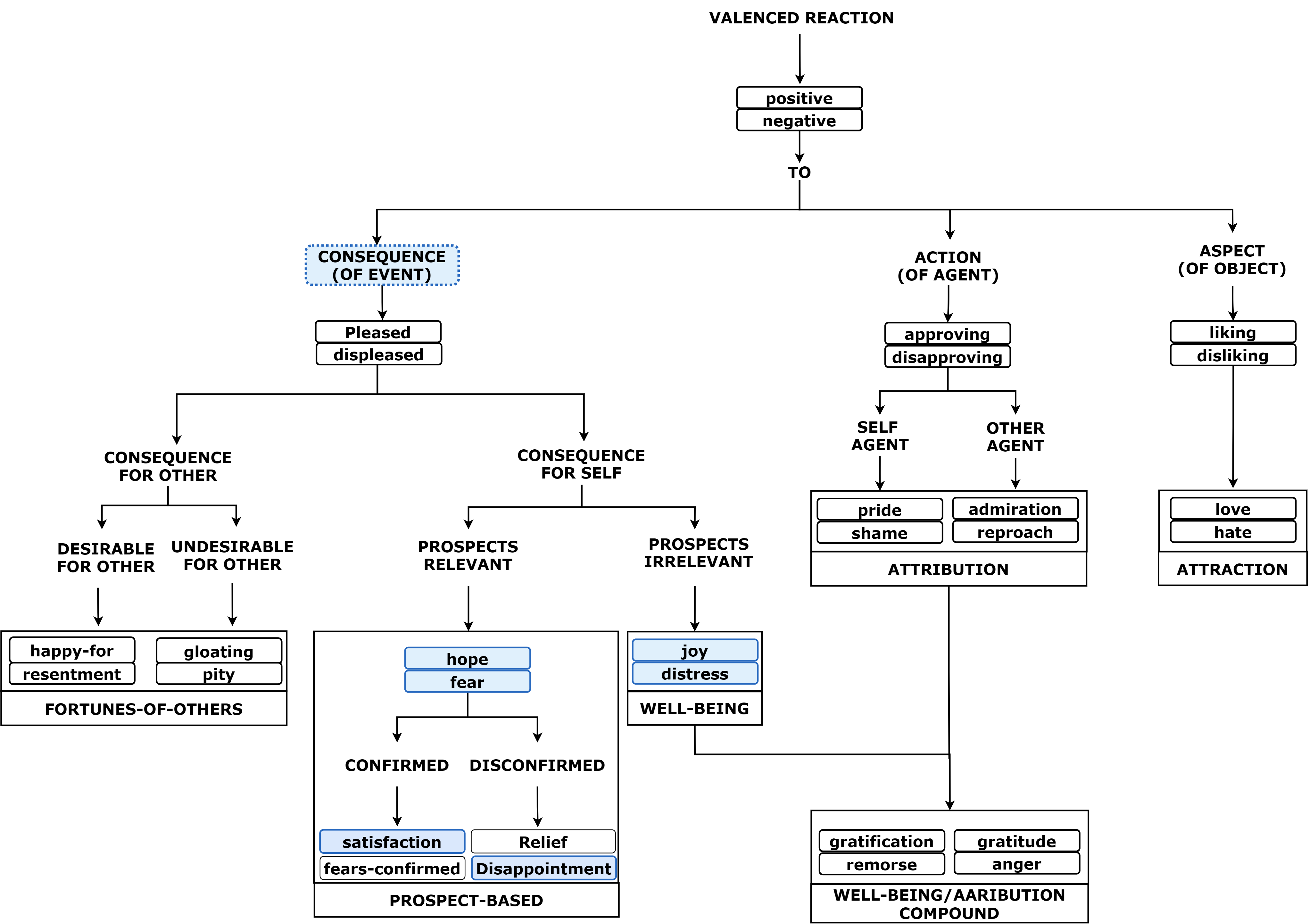}}
\caption{ OCC structure of emotions \cite{ortony1988cognitive}.}
\label{figOCC}
\end{figure}

Ortony, Clore, and Collins \cite{ortony1988cognitive} presented a cognitive structure of emotions which characterizes 22 emotion types (e.g. joy, hope, disappointment, distress and fear). According to their 'OCC' theory, emotions are valenced reactions which can be turned on by outcome of events, outcome of agents’ actions, or attributes of objects. 
Event-based emotions that are applicable to most game setups are highlighted in blue in Figure \ref{figOCC}. We selected them to be the basis of our proposed event-based transition system for emotions in our PX testing framework (further explanation in Section \ref{subsec.UXarchitecture}). Each of the emotion types listed in Figure \ref{figOCC} is specified as described in \cite{ortony1988cognitive}.

Table \ref{tblemotionspecifications} \HIDE{shows} summarizes OCC specifications of the highlighted emotion types; e.g. the OCC theory defines joy as {\em  is being pleased about a desirable consequence of event}. For example, consider a maze game in which an agent is looking for gold. When the agent finds a room with a gold pile, and it takes one step toward the gold, this has a desirable consequence (the agent is certain that it gets closer to the gold), so the agent feels pleased and as a result it starts to feel joy for the gold.
However, satisfaction is different. It is defined as
{\em being pleased about the confirmation of the prospect of a desirable consequence}. This emotion needs achievement confirmation whereas joy can be triggered whenever the agent becomes certain that the goal is achievable, although not fulfilled yet.
In the above example, satisfaction is triggered when the agent actually acquires the gold.
Additionally, while joy affects satisfaction, the agent might not be satisfied towards every goal which it is joyful about. In the earlier set-up,  the agent, when proceeding to collect the gold, faces guardians that need to be defeated first, and ends up consuming a unique item to win the combat. Thus, despite reaching the goal that it is joyful about, it would not be satisfied for failing to keep all its prized possessions.

\begin{table}
\vspace{-8mm}
\centering
\caption{Selected Emotions specifications according to the OCC theory \cite{ortony1988cognitive}.}\label{tblemotionspecifications}
\begin{tabular}{|rl|}  
\hline
Joy: & pleased about a desirable consequence of event \\ 
Distress: &displeased about an undesirable consequence of event \\
Hope: &pleased about the prospect of a desirable consequence of event \\
Fear: &displeased about the prospect of an undesirable consequence of event \\
Satisfaction: & pleased about the confirmation of the prospect of a desirable consequence \\
Disappointment: & displeased about the disconfirmation of the prospect of a desirable consequence \\
 \hline
\end{tabular}
\end{table}

\vspace{-3mm}
In general, dealing with emotions involves \textit{appraisal} and \textit{coping} \cite{ortony1988cognitive}. 
When an agent receives an event, the appraisal process is triggered to form emotions. Afterward, the agent responds to those emotions  based on coping strategies which affects the agent behavior towards the environment. In other words, emotions regulate the agent's actions during the coping process. In this paper, we focus on modeling of appraisal ---the proposed appraisal model of event-based emotions will be presented in Section \ref{sec.model}.

\section{ Agent-based Player Experience Testing Framework} \label{sec.UXframework}
In this section, we will explain the proposed framework architecture with their components and demonstrate appraisal in PX testing with some examples.

\subsection{The framework architecture } \label{subsec.UXarchitecture}

\begin{figure}[htbp]
\centerline{\includegraphics[width=80mm]{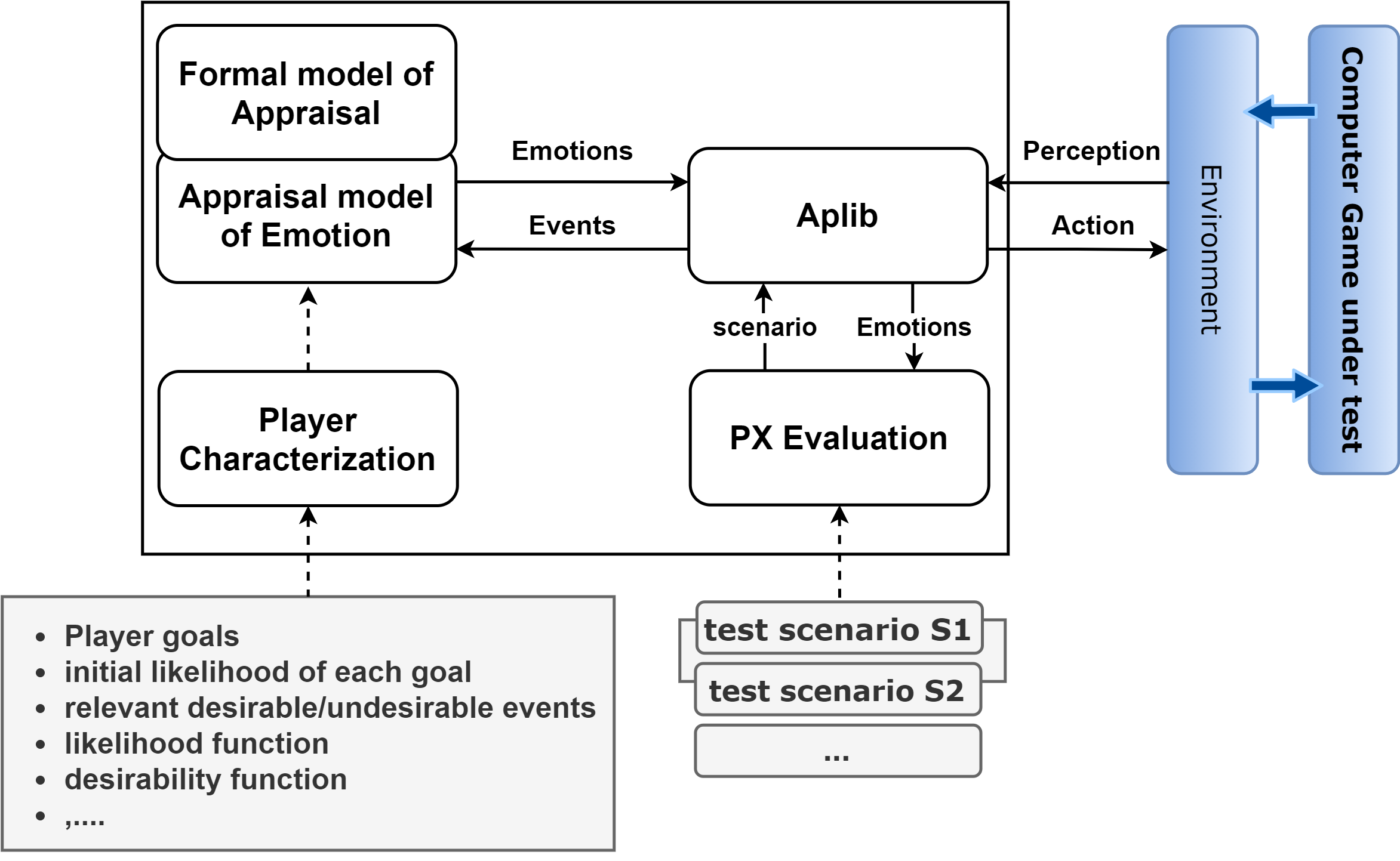}}
\caption{Automated PX testing framework architecture.}
\label{figframework}
\end{figure}

The general architecture of the proposed framework is presented in Figure \ref{figframework}, showing {\sf appraisal model of emotions}, {\sf player characterization}, \Aplib \ and {\sf PX evaluation} as the key components. They are defined below.

{\bf Appraisal model of emotions}. Emotions of a test agent are modeled based on the OCC theory. Since the framework does emotional evaluations on perceived events produced by the agent's actions or uncontrollable dynamic events such as hazards, only event-based emotions relevant to computer games are needed. 
To model these emotions\HIDE{according to OCC theory}, a {\em transition system approach} is proposed, which is formalized in Section \ref{sec.model}. This calculates the event-based emotion types with their respective intensity.
We will focus on a single test agent setup, thus for now we leave out emotions that are only valid in multi-agent settings. There is also room for extending the model, in the future, to test aspects of players' experience that are formed in various social contexts.

{\bf Player characterization}. Some properties of the appraisal model of emotions need to be specified by game designers with respect to the game under test as well as the player characteristics. For example, the designers should specify what goals are relevant for players (e.g. winning the game, collecting in-game money), what  in-game events are relevant to these goals, and in what way they are related to the goals (are they desired towards reaching a goal, or else undesirable?). Additionally, the desirability of an event might differ from one player character to another. Thus, player or set-up dependent properties must be initially set in this part of framework, before running the model of appraisal. 
Having such a component in our framework also provides an opportunity to enhance it in the future with more advanced characteristics such as players' moods and play-style (e.g., exploratory or aggressive \cite{stahlke2018usertesting,zhao2020winning}).

{\BfAplib}\footnote{https://iv4xr-project.github.io/aplib/} \cite{prasetya2020aplib}. A Java library for programming intelligent agents. It provides an embedded Domain Specific Language (DSL) to use all benefits of the Java programming language. \Aplib\  has a  BDI architecture \cite{herzig2017bdi} with a novel layer for tactical programming to control agents behavior more abstractly. Despite other use cases, the library has been developed for testing tasks in highly interactive software like games.

{\bf PX Evaluation}. Designers give test scenarios to the framework to check whether their newly developed content indeed triggers the expected emotions. This part is responsible for the visualization of the emotional state of the test agent as it pursues dedicated goals in a game environment with a given test scenario. Generated emotion types with their upward/downward trends during the test would assist designers to alter game parameters to optimize the experience in a certain degree.

\subsection{Appraisal theory in PX testing }
As mentioned earlier, the appraisal process is an essential part of  computational models of emotions. So, to automatically test the player experience based on emotions, we need to include this process in our framework for creating emotions. This would allow us to check whether the designers' expected emotions are  as same as the triggered players' emotions when  exposed to certain situations in the game.

\HIDE{Another example is assessing emotional engagement in games which is associated with player experience.} For instance, educational games are often evaluated based on the engagement level of learners to promote learning. Traditionally, to do this, players' emotions are tracked using either self-reports or automated facial emotion detection during a game-based task \cite{ninaus2019increased}, Identifying positive and negative  emotions plays an essential role in deciding if some game-based conditions and tasks need to be changed to optimize learning. Our proposed framework would help in performing this process automatically using  model of emotions to create emotions with respect to events.

Users of a more traditional, non-game, system typically need to feel higher levels of positive emotions and low levels of negative emotions to reach a satisfactory experience, while moderate levels of positive emotions and a high level of negative emotions such as distress, fear and disappointment could end up in an unsatisfactory experience with the system \cite{partala2012understanding}. These negative emotions reflect users' feelings when they are unable or unsure of how to use the system in some situations. This lead to the poor usability of the system \cite{saariluoma2014emotional}. 
However, computer games, e.g. those in the RPG and combat genres, can be deliberately designed to invoke certain negative emotions for certain experience in players because it can ultimately contribute to their enjoyment \cite{bopp2016negative} or even lead to high level of positive emotion when the player overcomes reasons that evoked negative emotions like fear and frustration \cite{lazzaro2009we}. 
Thus, unlike UX testing, in PX testing designers also need be able to \HIDE{be done in terms of} analyze relations between \HIDE{both} positive and negative emotions.
Our proposed framework can automatically check whether these emotions are appraised during playing the game. The prototype further refines this by also tracking when and where these emotions occur, thus enabling refined analyses. If the patterns of these emotions do not meet expectation, designers can change properties of the game and iterate the emotional testing process to achieve the expected emotions.

Ultimately, modelling a player's coping process improve the ability of the framework in PX testing. This is discussed briefly in Section \ref{sec.concl}. 
However, being able to model the coping behavior does not change the fact that the framework needs to also support the appraisal process of emotions in the first place. \HIDE{Therefore, our proposed framework bases on the appraisal process.}
For this reason, our proposed framework first focuses on the appraisal process.

\section{Event-based Formal Model of Emotion}
\label{sec.model}
Imagine that a software testing agent which takes the role of the player is deployed on a computer game to do PX testing. The agent is modelled as an event-based transition system which can appraise emotions to emulate the emotional state of a player. Its state consists of its 'belief' (perception) over the game and its emotions which can eventually affect its behavior to resemble the player behavior. In this section, we describe the essential part of the formalization of this event-based emotion transition system to conduct an approach for formal modeling of automated PX testing.

In the following, we assume an agent to have beliefs and goals, based on which it decides which actions should be taken in the environment. Being able to differentiate between different goals is useful for PX testing, as games often offer various optional plots and goals to players to improve their non-linearity and replay value. A goal $g$ is represented as a pair $\langle id, x \rangle $, with $id$ as its unique identifier and $x$ as its significance or priority of the goal. Goals and their significance are static in this setup.
We also assume that an agent senses its environment by means of events. For simplicity, it is assumed that the agent observes one event at a time, causing \HIDE{state transitions and moves} the agent to transition from one state to another. Whereas the agent's own actions are events, there are also events that arise from environmental dynamism such as hazards and updates by dynamic objects. We also add the event $tick$ to discretely represent the passing of time.
We represent emotion types as $Etype$\ = $\{\ Joy, Distress, Hope, Fear,...\}$. In the sequel, $etype$ ranges over this set.
\newtheorem{defn}{Definition}

\begin{figure}[htbp]
\vspace{-5mm}
\centerline{\includegraphics[width=6cm]{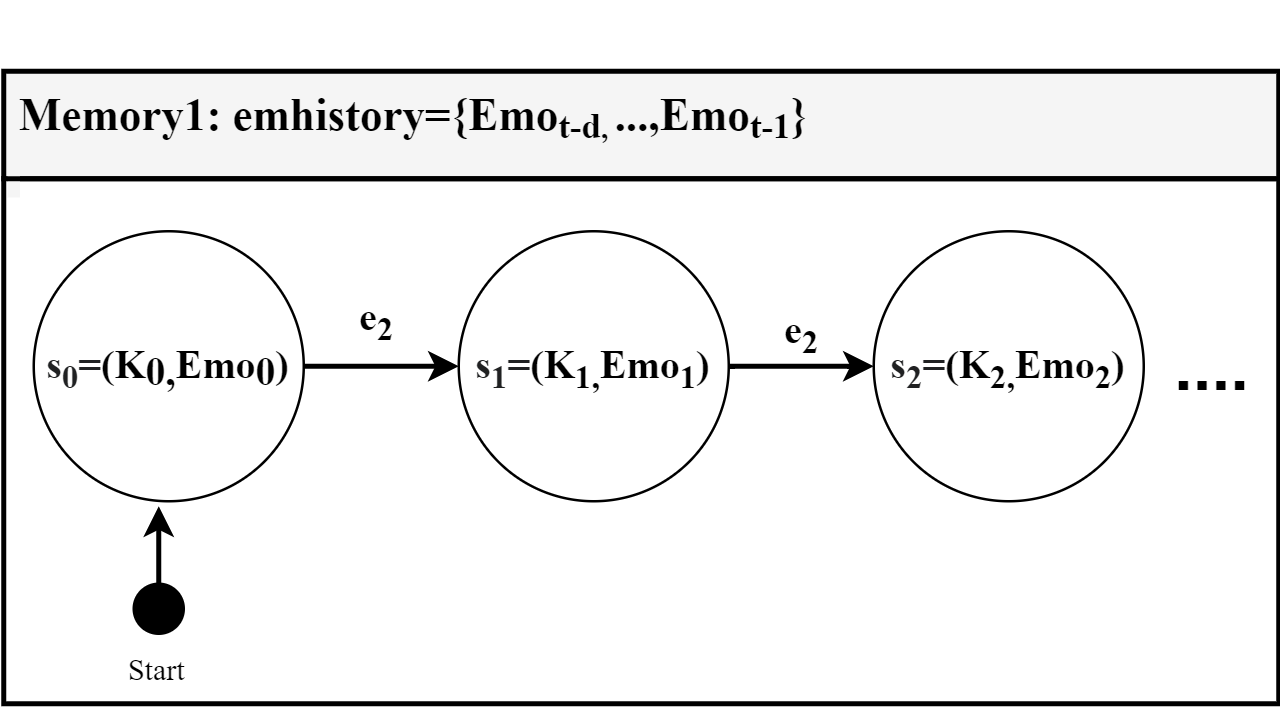}}
\caption{An agent's state transitions, as it receives an event $e_1$ followed by $e_2$.
\HIDE{Overview of state transitions}}
\label{figtransition}
\end{figure}

\begin{defn}\label{def.agent}
\normalfont
\label{one}
An emotional testing agent is represented by a transition system $M$, described by a tuple:  \[\langle\Sigma,s_0,G,E, \delta, \Pi,T\!hres\rangle
\]
%
where:
\begin{itemize}

    \item $G$ is a set of the agent's goals. 
    \item $\Sigma$ is the set of $M$'s possible states.
    Each state $s$ in $\Sigma$ is a pair $\langle K,Emo\rangle$ where:
    \begin{itemize}

        \item $K$ is a set of propositions representing the  agent's beliefs. 
        We additionally require that for every $g \in G$, $K$ 
        includes a proposition representing the goal's confirmation or dis-confirmation status, and a proposition representing the likelihood of reaching this goal from the current state. The former is represented by \HIDE{the proposition} $status(g,p)$ where $p \in \{achieved, failed, proceeding\}$ and the latter by $likelihood (g,v)$ where $v \in [0..1]$.
    
        \item  $Emo$ is a set containing the agent's active emotions, each is represented by a tuple $\langle etype, w , g,  t_0 \rangle$ specifying the emotion type $etype$, its intensity $w$ with respect to a goal $g$, and the time $t_0$ at when the emotion is triggered.

    \end{itemize}
    \item $s_0 \in \Sigma$ is  the initial state. It should specify the agent's initial belief on the likelihood of every goal, as well as initial prospect-based emotions (hope and fear). The rationale for the latter is that having an initial prospect towards a goal implies that there is also hope for achieving it, as well as some fear of its failure.
    \item  $E$ is the set of events the agent experiences.
    
    %

    \item  $\delta: \Sigma{\times}E \rightarrow \Sigma$ is the state transition function that describes how $M$ moves from one state to another upon perceiving an event. The definition is rather elaborate, and will be given separately in Definition \ref{def.delta}. 
   
   \item $\Pi=\langle Des, Praisew,DesOther, Liking \rangle$ is a tuple of appraisal dimensions according  to the OCC theory. This determines how an event is appraised in terms of its {\em desirability}, {\em praiseworthiness}, {\em desirability by others} and {\em liking}.
   
    
   \item $T\!hres$ is a set of thresholds, one for every type of emotion.
\end{itemize}

As an example, Figure \ref{figtransition} \HIDE{gives an overview of} illustrates first few transitions. We, additionally, assume the agent maintains an emotional memory, called $emhistory$, which keeps the history of active emotions ($Emo$) for a reasonable time window in the past: 
        \[ 
        emhistory \ \ = \ \ \overbrace{Emo_{t-d}
         \ ,\ \ ... \ \ ,\  Emo_{t-1}}^{\mbox{\footnotesize time window $d$}}
        \]
        where $t$ is the current system time and $d$ is the size of the memory's time window. $Emo_{t-i}$ indicates the active emotional at time $t-i$ in the past. 
        
\end{defn}

Before presenting the rest of the formal model, we feel the necessity to bring more clarity into the concept of goals' likelihood and status. The transition system is defined in a way that there is a slight difference between $likelihood(g,1)$ and $status(g,achieved)$. When an agent experiences  $likelihood(g,1)$, it is possible that the goal $g$ does not get confirmed in the same state. In other words, the agent comes to believe that the goal is reachable with 100\% certainty, but the achievement of the goal has not been confirmed yet in the current state. A similar relation holds for $likelihood(g,0)$ and $status(g,failed)$. 

The next key point 
is the agent's appraisal component $\Pi$, which has four dimensions. They help in modeling how events are appraised with respect to every goal in the corresponding dimension. Each appraisal dimension is described as a function over the agent's
beliefs, an event and a goal: $\Pi_\textbf{ \small Dim }(K, e,g), \mbox{where} \ Dim {\in} \{ Des, Praisew,DesOther, Liking\}$.
For example, $\Pi_\textbf{ \small Des}(K, e,g)$ determines the desirability of an event $e$ with respect to the goal $g$, judged when the agent believes $K$; the latter implies that this desirability might change when $K$ changes. Depending on the emotion, one or multiple appraisal dimensions might be triggered. Currently, $\Pi_\textbf{ \small Des}$ is the only dimension being actively used in our model because according to the OCC theory, the only appraisal dimension which affects our selected emotion types is the desirability function. However, we keep the structure in the general form for  possible future extension of the emotion types.

Below we will explain how emotions will be calculated, but importantly we should note that PX designers must provide some information as well, namely the following components of the tuple in Definition \ref{def.agent}: (1) the goal set $G$, along with the significance \HIDE{($significance(g,x)$)} and initial likelihood of each goal ($likelihood(g,v_{init})$), (2)  likelihood functions modelling how events affect the agent's belief towards goals' likelihood, (3) the appraisal dimensions, in particular $\Pi_\textbf{ \small Des}(K, e,g)$, (4) the thresholds \textit{Thres} and (5) decay rate $decay_{etype}$. 
In the simplest form, $\Pi_\textbf{ \small Des}(K, e,g)$ can be described by a mapping that maps events to the goals they are perceived as desirable/undesirable. In a more refined description this can be a function that monotonically increases with respect to the goal significance and likelihood\HIDE{over interval [-1,1]}. 
In terms of the architecture in Figure \ref{figframework}, the above
components are described in the {\em Player Characterization} part.
\HIDE{by PX designers according to the game under test.} 
\HIDE{PX designers determine the likelihood function of every goal over interval [0,1].} 


\begin{defn} Event-based Transition. \label{def.delta}
\normalfont
As mentioned earlier, the agent's state transition is driven by one incoming event  at a time. The transition function ($\delta$ in Def. \ref{def.agent}) is defined as follows. Let $e$ be an occurring event:
%
\[\langle K\ ,\  Emo\rangle \ \ \xrightarrow{\ \ \  e\ \ \ } \ \ \langle K'\ , \ \overbrace{newEmo(K,e,G) \ \oplus \ decayedEmo(Emo)}^{\mbox{\scriptsize updated emotion $Emo'$}} \rangle 
\]
where:

\begin{itemize}

     \item $K' = e(K) \setminus H$, where $e(K)$ is the agent's new beliefs obtained by updating $K$ with event $e$;
     here, the event $e$ is assumed to have a semantic interpretation as a function that affects $K$, including the parts that concern goals' likelihood and status.
     $H$ expresses likelihood information that can be removed from $e(K)$, because the corresponding goals are achieved or failed. More precisely, $H$ is the set
     $\{ \ likelihood(g,v) \ | \ status(g,p) \in e(K), \
     p\in\{achieved, failed\}, \ v \in \{0,1\}\; \}$.

    \item $Emo'= newEmo(K,e,G)  \oplus  decayedEmo(Emo)$ is the agent's emotions updated by the perceived event $e$ and the agent's new beliefs. Importantly, the $newEmo(K,e,G)$ specifies the {\em newly} triggered emotions 
    (see Def.\ref{def.emotype}), whereas
    $decayedEmo(Emo)$ (see Def.\ref{subsec.decay}) is a set of active emotions that decay over time. The operator $\oplus$ merges all these emotions after applying some constraints to have the updated emotional state of the agent. The emotional update is explained in Section \ref{subsec.emo.update}. 

\end{itemize}
\end{defn}  



When an agent perceives an event (except $tick$ event), new emotions may be triggered. This is done by calculating a so-called 'emotion function' $\mathcal{E}$ for every emotion type, as follows: 
 \[ \mathcal{E}_{etype}(K,e,g)  = w \]
This function specifies the activation intensity $w$ of the emotion $etype$ towards the goal $g$, as a consequence of the occurrence of $e$ and having beliefs $K$. Importantly, note that the function expresses {\em goal oriented} emotions, whereas the OCC theory includes e.g. emotions towards events or objects. We focus on goal oriented emotions due to the importance of goals, ranging from defeating monsters to getting the highest score, for game players.
A $tick$ event is used to represent the passing of time. This event would cause decays of active emotions in the transition system.
The definition of newly triggered emotion, mentioned in Def.\ref{def.delta}, is given below. It is used whenever a new emotion is triggered or an existing emotion reoccurs in the system. The way these new emotions are merged with existing emotions in $Emo$, as mentioned in Def.\ref{def.delta}, will be explained in Section \ref{subsec.emo.update}. We also need to remind that some hope and fear already exist in the system at the beginning which can be re-triggered by this function. Their initial values are set according to goals' significance and initial likelihoods of goals.

\begin{defn} \label{def.emotype}New Emotions.
\normalfont
The set of new emotions  triggered  by $e$ is:
\[ 
\begin{array}{l}
 new Emo(K, e, G) = \   \{\langle etype,g,w,t\rangle\ \ | \ etype \in Etype, \ g \in G,\;
 w=\mathcal{E}_{ etype}(K, e,g)>0 \}
\end{array}
\]
where $t$ is the current system time that the emotion is triggered.

\end{defn}
In the above definiton $E_{etype}$ is a so-called activation emotion function that calculates the activation intensity for different newly triggered event-based emotion types. Each activation emotion function has an activation potential and a threshold which form the activation intensity of the newly triggered emotion (see Def.\ref{def.joy}). The level of desirability an event respecting a goal and the agent's goal likelihood are the main variables affecting the activation potential as hinted in the OCC theory. To trigger a new emotion type, its activation potential value needs to pass the corresponding threshold. 
The concept of threshold is needed if we want to \HIDE{have experiments} support setups with different agent's moods because the thresholds depend on the moods
(e.g. Steunebrink et al.\cite{steunebrink2007logic} pointed out that with a good mood, the thresholds of negative emotions increase, hence bringing about a lower degree of intensity in negative emotions when they are triggered). 
All activation functions of emotions defined below have the same structure. However, the potential part might differ. They are as follows\footnote{For convenience, we only define the functions partially. The cases where they are undefined will be ignored by Def. \ref{def.emotype} anyway, where they are used.}:

\begin{defn}Joy \label{def.joy}
\vspace{-3mm}
\[
  \mathcal{E}_\textbf{ \small \emph{Joy}} (K,e,g)  \ \ \ = \ \ \ 
  \overbrace{
  \underbrace{\Pi_\textbf{ \small Des}(K,e,g)}_{\mbox{\scriptsize activation potential}} 
  \ \ - \ \ Thres(Joy)}^{\mbox{\scriptsize activation intentsity}}
\]
\normalfont
provided $g \in G$,\;
  $likelihood(g,1) \in e(K)$\footnote{\small Unlike prospect-based emotions, well-being emotions are certain. So, joy and distress towards a goal only happen if the goal's likelihood becomes 1 and 0 respectively. In particular, obtaining certainty of achieving/failing the goal is seen as notable desirable/undesirable consequence of an event to justify these emotions. There might other practical consequences, but we will mostly focus on the aforementioned types of consequences.
  },
   and $\Pi_\textbf{ \small Des}(K, e,g) > 0$.

\end{defn}

\begin{defn}Distress
\[
  \mathcal{E}_\textbf{ \small \emph{Distress}} (K,e,g)  \ = \ \lvert \Pi_\textbf{ \small Des}(K,e,g) \rvert - Thres(Distress)\]
\normalfont
provided
  $g \in G$,
  $likelihood(g,0) \in e(K)$,
   and $\Pi_\textbf{ \small Des}(K, e,g) < 0$.
   Unlike $Joy$, $Distress$ is triggered when an event is \HIDE{evaluated} deemed as undesirable towards the goal.
\end{defn}

\begin{defn}Hope
\[
  \mathcal{E}_\textbf{ \small \emph{Hope}} (K,e,g)  \ = \ v'*x-Thres(Hope) \]
\normalfont
provided 
   $g=\langle id,x \rangle \in G$, $likelihood(g,v) \in K$,  
  $likelihood(g,v') \in e(K)$,
  and $v{<}v'{<}1$.

  It is assumed that the increase in likelihood of a goal is only possible if the incoming event is desirable towards the goal. Thus, with this assumptions, there is no need to check the desirability of the event $\Pi_\textbf{ \small Des}(K,e,g)$ for prospect-based emotions.
\end{defn}

\begin{defn}Fear
\[
  \mathcal{E}_\textbf{ \small \emph{Fear}} (K,e,g)  \ = \ (1-v')*x-Thres(Fear) \]
\normalfont
provided $g=\langle id,x \rangle \in G$, $likelihood(g,v) \in K$, 
$likelihood(g,v') \in e(K)$, and $0{<}v'{<}v$.

\end{defn}

\noindent

\begin{defn}Satisfaction

\[
  \mathcal{E}_\textbf{ \small \emph{Satisfaction}} (K,e,g)  \ = \ x-Thres(Satisfaction) \]
\normalfont
provided 
  $g{=}\langle id,x \rangle \in G$, $status(g,achieved) \in e(K)$, 
 $\langle Hope,g \rangle \in emhistory$, and $\langle Joy,g\rangle \Inn emhistory$.
\end{defn}
\noindent

\begin{defn}Disappointment
\[
  \mathcal{E}_\textbf{ \small \emph{Disappointment}} (K,e,g)  \ = \ x-Thres(Disappointment) \]
\normalfont
provided 
 $g{=}\langle id,x \rangle \in G$, $status(g,failed) \in e(K)$, 
 $\langle Hope,g \rangle \in emhistory$, and $\langle Distress,g\rangle \Inn emhistory$.

\end{defn}

\subsection{Decay of emotions} \label{subsec.decay}

Every emotion  has a duration called \textit{emotion episode} in which the peak of its intensity, its decay rate, possible recurrences, and the time that the emotion is triggered are shown\cite{steunebrink2007logic}.
As indicated earlier in Def.\ref{def.agent}, $tick$ is a time event to show the passing of time in our transition system. We can reflect decays of emotions using this event:
{\small
\[ \langle K, Emo\rangle {\xrightarrow{\ \ \  e=tick \ \ \ }}\langle K', Emo'\rangle\]
}
where $K'$ and $Emo'$ refer to the updated beliefs and updated active emotions after the transition.
The intensity of active emotions in $Emo$ would decrease 
as follows:
%
{\small
\[ \begin{array}{l}
      decayedEmo(Emo)
      \ = \\
      \hspace{5mm}\{\; \langle etype,g,w',t_0 \rangle \ \ | \ \ \begin{array}[t]{l}\langle etype,g,w,t_0\rangle \in Emo, \ \ w'={\sf intensitydecay}_\textbf{ \small etype}(w_0,t_0)>0, \\ w_0=emhistory(etype,g,t_0) \; \}
      \end{array}
\end{array}
\]
}
where $w_0=emhistory(etype,g,t_0)$ denotes the initial intensity of $etype$ with respect to $g$ which can be obtained from $emhistory$.
  There is not a unique quantitative formalization for the decay function $\sf intensitydecay$. This function can be defined in a way which relates the usage and the interpretation of decay \cite{steunebrink2008formal} \cite{dias2014fatima}. However the peak of intensity ($w_0$), the time at which the emotion is  triggered ($t_0$) and the decay rate ($decay_{etype}$) are essential parameters that must be taken into account. While an inverse sigmoid decay function is proposed by \cite{steunebrink2008formal} to reflect the gradual decrease of intensities\HIDE{in the system}, \cite{dias2014fatima} is making use of a negative exponential function with almost the ame parameters. We used the latter decay function \cite{dias2014fatima} in our model although the sigmoid decay function \cite{steunebrink2008formal} can be used as well.
  \[{\sf intensitydecay}_\textbf{ \small etype}(w_0,t_0)=w_0  \ * \ e^{\ c \ * \ decay_{etype} \ * \ (t-t_0)} \ \ \ , -1<c<0\]
where \ $t$ \ is \ the \ current \ system \ time \ and \ $t_0$  is the \ time \ at \ which \ the \ emotion \ starts.

\subsection{Inconsistent emotions}\label{subsec.axioms}
Emotions are triggered regarding the goals, so technically the agent might have several emotions towards the same goal. Nevertheless, the OCC theory states that some emotions are mutually exclusive which means a human can not have them simultaneously for the same goal \cite{steunebrink2007logic}. These mutual exclusions, which should then also be held in every state of our transition system, are as follows:
%

%
\[
\begin{array}{lcl}
Emo' \ \vDash \ \neg{(  \langle Hope,g \rangle \wedge  \;  \langle Joy,g \rangle )} \\
Emo' \ \vDash \ \neg{(  \langle Fear,g \rangle \wedge  \; \langle Distress,g \rangle  )}
\end{array}
\]
As it is explained in Section \ref{sec.preliminaries}, whereas emotions such as hope and fear are prospect-based emotions which means they are uncertain ($likelihood(g,v)$), emotions like joy and distress are certain \cite{steunebrink2007logic}, so it is illogical to have both in the system. For example, when a player is joyful of acquiring the key to an in-game treasure room, because now the treasure should certainly be within his/her reach, this joy would now replace what was merely hope for getting the treasure.
In general, in case of happening a certain emotion, it replaces the corresponding prospect-based emotion, so the mutual exclusions are always maintained. We formulated our formal model in a way that in case of the conflicting emotions, the new certain emotion would take the place of the prospect-based emotion. However, the set of inconsistent emotions can be expanded based on the test purpose or the game under test. The designer can specify these as assumptions in the {\em Player Characterization} component. A notation as $axiomset(\langle etype,g \rangle)$ is used to access every rule containing $\langle etype,g \rangle$. 

\subsection{Emotional state update } 
\label{subsec.emo.update}

To update the emotional state, newly triggered emotions, $newEmo$, need to be merged with existing active emotions whose intensities are decreasing gradually, $decayedEmo$, to yield the new emotional state $Emo'$. There are three cases to consider.
Case-1 involves existing emotion types that decay without having the same emotion type or the conflicting type in the $newEmo$; these will be kept.
Case-2 involves newly triggered emotion types that do not exist in $decayedEmo$; these are added to $Emo'$.
Case-3 involves emotion types in $decayedEmo$ that reoccurs in $newEmo$. Only emotions from these three cases will be included in $Emo'$. In particular, this implies that in the cases of inconsistent emotions, the newly triggered emotion takes precedence over the emotion which has already existed \HIDE{in the emotion set to}by taking its place in order to uphold the mutual exclusions discussed before. The new one is added to $Emo'$ based on Case-2. This comes from the rationale that new belief and perceptions convey more accurate information than past information, and therefore the triggered new emotions have more weight for the player. The last case, Case-3, is about existing emotions that get {\em re-stimulated by} the new perceived event. To date there is no definitive answer to the question of how this should be reflected to the intensity of the corresponding emotions. We decided to take the maximum intensity value of the emotion (the dominant value). However, a more proper answer to the question would need further research. 
The update is formally shown below, with the Cases indicated accordingly: 
%
{\small
\[
Emo' = 
   \left\{
   \begin{array}{l}
   \mbox{\textcircled{\scriptsize 1}} \ 
   \{\langle etype,g,w,t_0\rangle \ | \ 
      \begin{array}[t]{l}\langle etype,g,w,t_0\rangle \in decayedEmo \\ 
      \wedge \ \neg \exists \  w', t'_0.\;  \langle etype,g,w',t'_0\rangle \in newEmo  \\ 
      \wedge \ 
      \neg \exists \  w',t'_0.\;
      \langle \overline{etype},g,w',t'_0\rangle \in newEmo  \ \}
      \end{array}\\ 
   \cup \\ 
   \mbox{\textcircled{\scriptsize 2}} \ 
   \{\langle etype,g,w',t'_0\rangle\ | \ 
      \begin{array}[t]{l}
      \langle etype,g,w',t'_0\rangle \in newEmo \\ \wedge \ \neg \exists w,t_0.
      \langle etype,g,w,t_0\rangle \in decayedEmo \ \}
      \end{array} \\
   \cup \\ 
   \mbox{\textcircled{\scriptsize 3}} \ 
   \{ {\sf max} (\langle etype,g,w,t_0\rangle , \langle etype,g,w',t'_0\rangle) | \ 
      \begin{array}[t]{l}
      \langle etype,g,w,t_0\rangle \in decayedEmo \\ \wedge \  \langle etype,g,w',t'_0\rangle \in newEmo\}
      \end{array} 
   \end{array}
   \right.
\]
}
 where $t_0$ is the time at which an emotion is triggered (starts)  and the outcome of ${\sf max}$ is the one with the higher intensity. An emotion that is in conflict with $etype$ is referred as $\overline{etype}$. The above update scheme will uphold the axiom
 $\neg (\langle etype,g\rangle \wedge \langle\overline{etype},g\rangle) \in axiomset(\langle etype,g\rangle)$. 

\comment{\subsection{\bf Goal chain}
\label{subsec.goalchain}
As indicated earlier, the beliefs $K$ gets updated according to the new event. In particular, this might affect the agent's belief towards the likelihood of achieving certain goals. Recall that this is modelled in the {\em Player Characterization} component in our approach, e.g. by means of some update rules.
However, modern games often offer multiple goals that players can go after, and furthermore have dependency. E.g. obtaining a unique item Excalibur might be an optional goal in a game, but achieving this might improve the likelihood of defeating the end boss. To capture this, we can extend the Player Characterization with 
'chained' rules, for example $R = \{e_1\rightarrow g_1,\; g_1\rightarrow g_2,\; g_2 \rightarrow g_3\}$ to express that the event $e_1$ affects the likelihood or status of the goal $g_1$, which in turn affects the likelihood of $g_2$ and so on. We do not write down how exactly the likelihood should be adjusted, but imagine that the rules also specify this.
When the agent received $e_1$, it should now not only apply the rule/update $e_1 \rightarrow g_1$, but also other rules in $R$ whose antecedent is transitively triggered by $e_1 \rightarrow g_1$.
While the rules in $R$ above can indeed be equivalently described by more direct rules of the form $\{e_1\rightarrow g_1,\; e_1\rightarrow g_2,\; e_1 \rightarrow g_3\}$, the chained form arguably captures inter-goal dependency more intuitively.}

\section{Proof of Concept}\label{sec.UXresult}

We conducted our experiment on a game called Lab Recruits\footnote{\url{https://github.com/iv4xr-project/iv4xrDemo/tree/occDemoPrototype}} which we subject to the combination of \aplib \ and our implemented model of appraisal\footnote{\url{https://github.com/iv4xr-project/jocc}} to provide the proof of concept and show our early results in PX testing. 
Lab Recruits is a 3D game developed in Unity which has different replayable levels. Each level is a laboratory building with a number of rooms containing interactable objects, such as button and non-interactable objects, such as desk and fire hazards.

\begin{figure}[htbp]
   \captionsetup{justification=centering}
    \begin{minipage}{0.5\textwidth}
        \subfloat[The floor plan of the level.]{\includegraphics[width=0.9\textwidth]{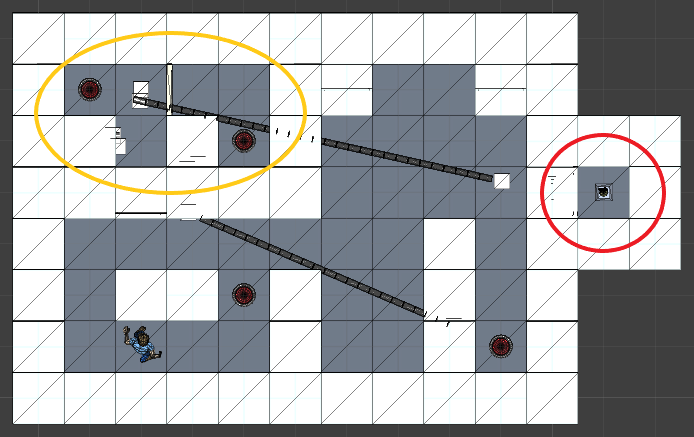}\label{subfigfloorplan}}
    \end{minipage}
    \begin{minipage}{0.5\textwidth}
    \subfloat[Setup 1.]{\includegraphics[width=0.5\textwidth]{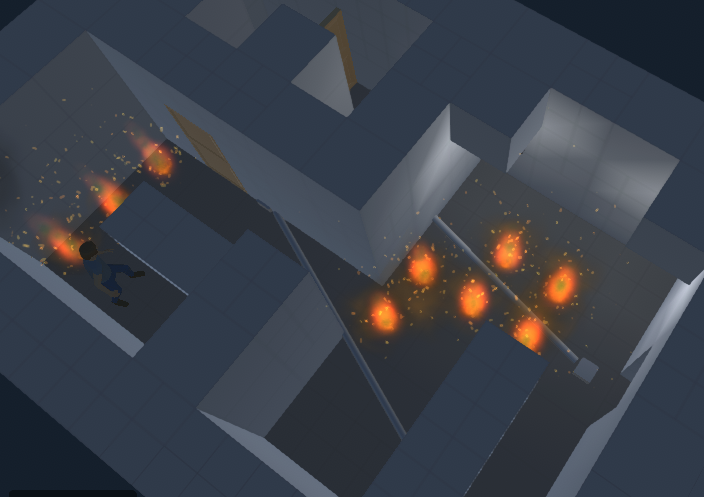}\label{subfigsetup1}}
    \subfloat[Setup 2.]{\includegraphics[width=0.5\textwidth]{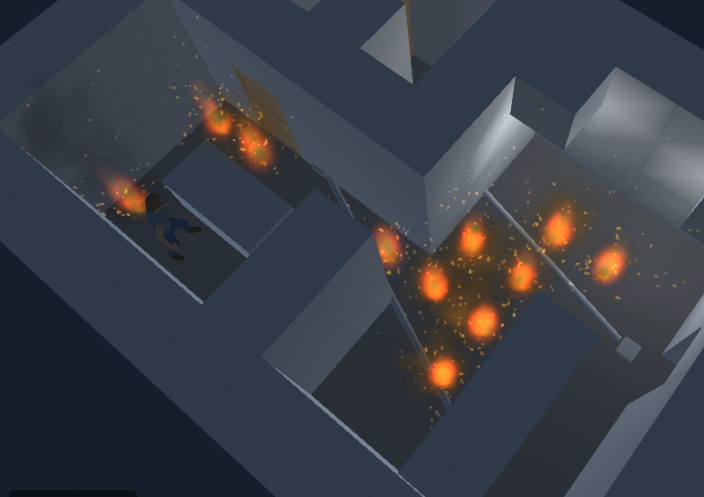}\label{subfigsetup2}}%
    \end{minipage}
    \caption{The level under the PX test in Lab Recruits.}%
    \label{figtestlevel}%
\end{figure}

Figure \ref{subfigfloorplan} shows the floor plan of the level exposed to PX testing using our approach. It consists of four buttons, three doors, and some fire hazards. The goal is for the player to escape the level by reaching the exit room circled in red. Access to this room is guarded by a closed 'final door'.
The level contains some rooms with a puzzle (yellow circle) that involves finding the buttons to open the final door and reopen the doors that in the process become closed to entrap the agent. Figure \ref{subfigsetup1} and \ref{subfigsetup2} show two provided setups with the different amount and locations of fire hazards. The agent will lose health points by passing each fire hazard.
These setups are examples of 
choices considered by designers, although being currently simple, as to which one would lead to better PX.

As mentioned in Section \ref{sec.model}, a developer sets needed inputs of the model such as the goal set, initial likelihood of each goal, the desirability of events for each goal, the threshold and decay rate of emotions in {\em Player Characterization}. A test agent is deployed, set with multiple goals, though here we will only discuss the most significant one, namely completing the level.
Initially, the agent is assumed to believe that the likelihood of achieving this goal is  0.5.
The agent is given a program so that it can automatically explore the level.
As the agent progresses, its belief on the likelihood
of completing the level changes, depending on the number of opened door as well as remained closed doors. Opening each door is assumed to have a desirable consequence for the agent because it increases the chance of the agent to complete the level.

\begin{figure}[htbp]
   \captionsetup{justification=centering}
    \centering
    \subfloat[Setup 1]{{\includegraphics[width=5.6cm, height=3.9cm]{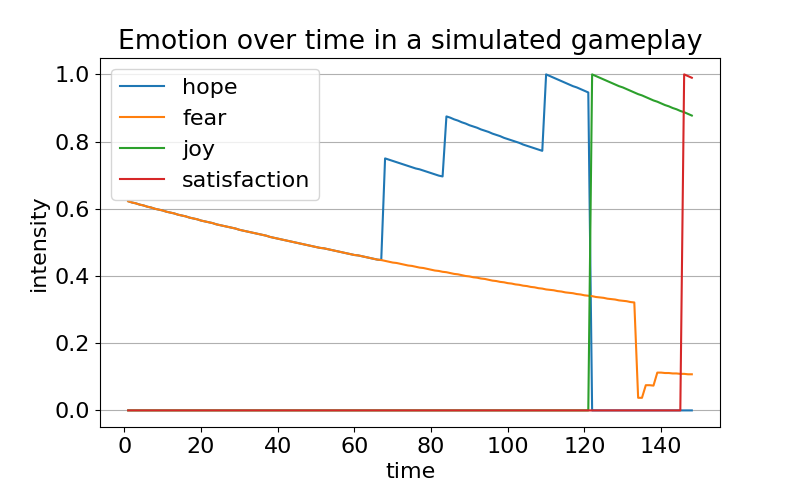} } \label{subfigtimeline.1}}
    \qquad
    \subfloat[Setup 2]  {{\includegraphics[width=5.6cm, height=3.9cm]{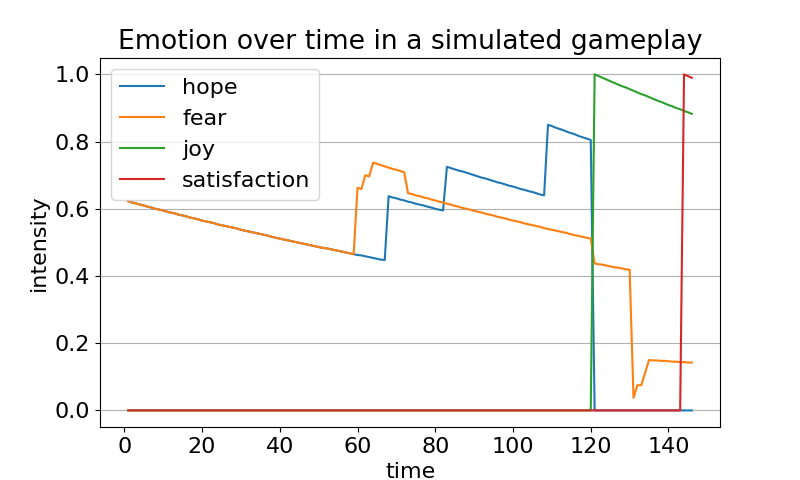} }\label{subfigtimeline.2}}%
    \caption{The emotions' timelines correspond two setups of the game level in Figure. \ref{figtestlevel}.\\
    {\small  (threshold=0, decay rate=0.005)} }%
    \label{figtimeline}%
\end{figure}

The timeline of triggered emotions in the agent with respect to the goal "completing the level" is shown in Figure \ref{figtimeline}, along with their intensity levels at each time. The agent initially experience some hope and fear due to the assumed initial belief that completing the game is possible, with the likelihood 0.5. 
In both setups, when the agent pushes the button that opens the first door (time=70\footnote{The system is event driven, so only events can change the likelihoods. All emotions decay until an event is perceived. However, we can add an event type to the system to decay the likelihoods when there is no event for some period of time to update the emotional state.}), the agent's hope regarding completing the game starts to increase.  It decays or gets re-stimulated according to the events  until time 120 when it is replaced by joy. The agent feels a level of satisfaction,  When completing the game.
Comparing two setups reveals something interesting. Fear shows a quite different trend in setup 2 (Figure \ref{subfigtimeline.2}). It is strongly stimulated multiple times during the execution, whereas the same emotion is rarely stimulated in setup 1, and when it happens, it happens with much less intensity (Figure \ref{subfigtimeline.1}), towards the end of the play, where it matters less.
Such comparison can be useful for designers e.g. to determine the amount, and placement, of hazards to induce certain degree of fear along with keeping the chance for satisfactory experience of accomplishing the goal. In our case, setup 1 is less likely to thrill the player, whereas setup 2 has a better balance of the quantity and placement of the fire, by generating fear, while still keeping the level survivable.

\begin{figure}[htbp]
    \captionsetup{justification=centering}
    \centering
    \subfloat[Negative emotions: yellow= high fear, Orange shades=low fear.]{{\includegraphics[width=5.6cm]{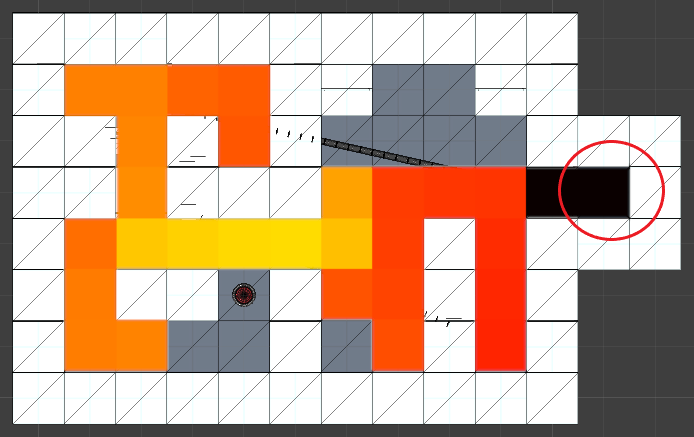} } \label{subfiggeat.1}}
    \qquad
    \subfloat[Positive emotions: Mahogany red=hope, Ruby red= joy, yellow= satisfaction.]  {{\includegraphics[width=5.6cm]{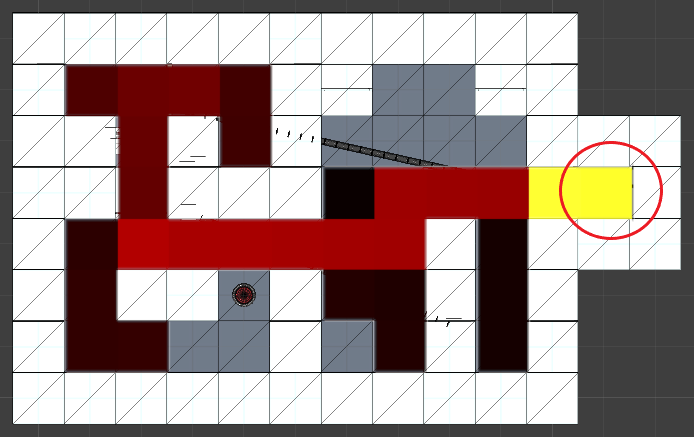}}\label{subfigheat.2}}%
    \caption{The heat maps of triggered emotions in setup 2.
    \hspace{\textwidth}
    Black=no emotion, white= walls (not walkable), gray=unexplored area.}
    \label{figheatmap}%
\end{figure}

 Figure \ref{figheatmap} shows some heat maps, providing spatial information of the agent's emotions in Setup 2. Comparing the outcomes of Figures \ref{subfigtimeline.2} and  \ref{subfiggeat.1} illustrates that the highest level of fear is experienced between time 65 to 75 when the agent is in a particular fire covered corridor (yellow in Figure \ref{subfiggeat.1}). Fire intensifies the agent's fear of failure, and moreover the agent has to walk this corridor several times. The most drastic decline in fear is when the agent is about to finish the level.

As can be seen in  Figure \ref{subfigheat.2}, the agent feels a higher level of hope when progressing in solving the buttons-doors puzzle in the puzzle rooms. After pushing the button that corresponds to the final door and reopening the door of puzzle room to escape it, the agent becomes certain that passing the final door is achievable now. Thus, the hope suddenly is replaced by the joy for reaching the final door to complete the game. At the end, the agent feels satisfied when the achievement is confirmed.
Having such information would help Lab Recruits designers to adjust the puzzles and fire hazards in such a way to induce certain emotions, at the right moments and the right places, which ultimately affect a certain aspect of player experience like enjoyment.

\section{Related Work} \label{sec.relatedwork}
PX researchers aim to understand the gaming experience to ultimately induce certain experience. Fernandez \cite{fernandez2008fun} outlines the influence of players'  emotional reactions and their profile in enjoyment by extending the usability methods to uncover relationships between game components and the degree of fun in players. Sanchez et.al \cite{sanchez2012playability} explained that usability of games can be defined in the term of {\em playablity}. They present a framework guided by attributes and properties of playability to characterise experience for PX evaluation and observing the relation between the experience and the developed elements of a commercial video game.
Psycho-physiological methods is among techniques to measure aspects of PX like flow and immersion. Jennett’s et al. \cite{jennett2008measuring} tries to develop a subjective and objective measure for immersion using questionnaires and eye movement tracking respectively. Drachen et al. \cite{drachen2010correlation} report a significant the correlation between heart rate, electrodermal activity and the self-reported experience of players in first-person shooter games. 
Zook and Riedl \cite{zook2012temporal} introduce a temporal data-driven model to to predict the impact of game difficulty to player experience. Results of their empirical study on a role-player combat game show the game, that tailors its difficulty to fit a player abilities, improves the player experience. Most of PX prediction techniques are data-driven which involve human players in the process and as a result, they demand a high level of human labor. This led researchers to investigate model-driven approaches.
A computational model of motivation is presented in \cite{guckelsberger2017predicting} to predict PX without the need of human player using empowerment, the degree of control an agent has over the game. The study measures empowerment by intelligent agents  to create levels with defined empowerment to induce different PX. This would help to produce desired content characteristics during the  procedural content generation.

Despite existing research on modeling the OCC theory, the theory has not been employed in the context of PX testing. Having a proper formalization of emotion would act as a bridge from psychological description of emotions to computational models of emotions which are translatable to codes. Formalization of emotions has been mostly done in the form of BDI logic.
Steunebrink \cite{steunebrink2007logic} deployed a formal model inspired by the OCC theory to specify the influence of emotions, specifically hope and fear, on a BDI agent's decisions. Later, a full version of the model with all 22 emotions is explained in \cite{steunebrink2008formal}. 
Dias et al.\cite{dias2014fatima} presents an OCC-based appraisal engine called FAtiMA (\textbf{F}earnot \textbf{A}ffec\textbf{TI}ve \textbf{M}ind \textbf{A}rchitecture) for creating autonomous agent characters that can appraise events and behave based on socio-emotional skills. Its main use case is to automate virtual characters in conversing with humans.
 FAtiMA is claimed to be inspired by the OCC theory to simulate emotional skills in autonomous agents. However, so far, no formal model has been introduced to evaluate the toolkit regarding the OCC theory. A BDI-like probabilistic formalization is described in  \cite{gluz2017probabilistic} for OCC event-based emotions during the appraisal. The study evaluates the desirability of consequences of an event based on the agent's goal and the degree that the consequence can improve the possibility of the goal achievement. Unlike other formalisations that give a high level function for appraisal variables, it proposes a more refined logic-base calculation for these variables and also tries to formalize 'effort' and 'realization' that are involved in appraising some event-base emotions.

\section{Conclusion \& Future Work} \label{sec.concl}

This paper presented an automated PX testing approach using an emotional model. An event-based transition system is introduced to model the appraisal for 
event-based emotions according to the OCC theory which is then combined to a Java library for tactical agent programming called \aplib \ to create an agent-based PX testing framework. Early results of our experiment with the prototype show that such a framework that can emulate players' emotions would let developers to investigate how emotions of players would evolve in the game during the development stage. By providing e.g. heat-map visualisations of triggered emotions and their timelines, designers gain insight on how to alter parameters of their systems to evoke certain emotions.

We are currently doing more advanced experiments using the case study, Lab Recruits, to investigate initial moods, emotions and their effect on certain aspects of PX as a future work. There are also some concepts like emotional intensity after a recurrence that are described with high level functions in the literature which need a calculation mechanism. In particular, we want to do further research on how exactly an emotion should regain its intensity level after a re-stimulation.
Furthermore, the proposed framework, if enhanced by the coping process, would be able to simulate the effect of emotions on players' behavior for further PX evaluations. However, this needs extension in our event-based transition system to support the coping process formally respecting 
the OCC theory.
We ultimately plan to conduct research on validation of our model by comparing our results with the data of human players.

\bibliographystyle{splncs04}
\bibliography{Bibliography}

\begin{thebibliography}{10}
\providecommand{\url}[1]{\texttt{#1}}
\providecommand{\urlprefix}{URL }
\providecommand{\doi}[1]{https://doi.org/#1}

\bibitem{adam2009logical}
Adam, C., Herzig, A., Longin, D.: A logical formalization of the occ theory of
  emotions. Synthese  \textbf{168}(2),  201--248 (2009)

\bibitem{alves2014state}
Alves, R., Valente, P., Nunes, N.J.: The state of user experience evaluation
  practice. In: Proceedings of the 8th Nordic Conference on Human-Computer
  Interaction: Fun, Fast, Foundational. pp. 93--102 (2014)

\bibitem{baier2008principles}
Baier, C., Katoen, J.P.: Principles of model checking. MIT press (2008)

\bibitem{bernhaupt2015game}
Bernhaupt, R.: Game user experience evaluation. Springer (2015)

\bibitem{bopp2016negative}
Bopp, J.A., Mekler, E.D., Opwis, K.: Negative emotion, positive experience?
  emotionally moving moments in digital games. In: Proceedings of the 2016 CHI
  Conference on Human Factors in Computing Systems. pp. 2996--3006 (2016)

\bibitem{dias2014fatima}
Dias, J., Mascarenhas, S., Paiva, A.: Fatima modular: Towards an agent
  architecture with a generic appraisal framework. In: Emotion modeling, pp.
  44--56. Springer (2014)

\bibitem{drachen2010correlation}
Drachen, A., Nacke, L.E., Yannakakis, G., Pedersen, A.L.: Correlation between
  heart rate, electrodermal activity and player experience in first-person
  shooter games. In: Proceedings of the 5th ACM SIGGRAPH Symposium on Video
  Games. pp. 49--54 (2010)

\bibitem{fang2010development}
Fang, X., Chan, S., Brzezinski, J., Nair, C.: Development of an instrument to
  measure enjoyment of computer game play. INTL. Journal of human--computer
  interaction  \textbf{26}(9),  868--886 (2010)

\bibitem{fernandez2008fun}
Fernandez, A.: Fun experience with digital games: a model proposition.
  Extending experiences: Structure, analysis and design of computer game player
  experience pp. 181--190 (2008)

\bibitem{gluz2017probabilistic}
Gluz, J., Jaques, P.A.: A probabilistic formalization of the appraisal for the
  occ event-based emotions. Journal of Artificial Intelligence Research
  \textbf{58},  627--664 (2017)

\bibitem{guckelsberger2017predicting}
Guckelsberger, C., Salge, C., Gow, J., Cairns, P.: Predicting player experience
  without the player. an exploratory study. In: Proceedings of the Annual
  Symposium on Computer-Human Interaction in Play. pp. 305--315 (2017)

\bibitem{herzig2017bdi}
Herzig, A., Lorini, E., Perrussel, L., Xiao, Z.: {BDI} logics for {BDI}
  architectures: old problems, new perspectives. KI-K\"{u}nstliche Intelligenz
  \textbf{31}(1) (2017)

\bibitem{jennett2008measuring}
Jennett, C., Cox, A.L., Cairns, P., Dhoparee, S., Epps, A., Tijs, T., Walton,
  A.: Measuring and defining the experience of immersion in games.
  International journal of human-computer studies  \textbf{66}(9),  641--661
  (2008)

\bibitem{lazzaro2009we}
Lazzaro, N.: Why we play: affect and the fun of games. Human-computer
  interaction: Designing for diverse users and domains  \textbf{155},  679--700
  (2009)

\bibitem{meyer2015bdi}
Meyer, J.J., Broersen, J., Herzig, A.: Handbook of Logics of Knowledge and
  Belief, chap. {BDI} logics. College Publications (2015)

\bibitem{ninaus2019increased}
Ninaus, M., Greipl, S., Kiili, K., Lindstedt, A., Huber, S., Klein, E.,
  Karnath, H.O., Moeller, K.: Increased emotional engagement in game-based
  learning--a machine learning approach on facial emotion detection data.
  Computers \& Education  \textbf{142},  103641 (2019)

\bibitem{ortony1988cognitive}
Ortony, A., Clore, G., Collins, A.: The cognitive structure of emotions. cam
  (bridge university press. Cambridge, England  (1988)

\bibitem{partala2012understanding}
Partala, T., Kallinen, A.: Understanding the most satisfying and unsatisfying
  user experiences: Emotions, psychological needs, and context. Interacting
  with computers  \textbf{24}(1),  25--34 (2012)

\bibitem{peterson2017understanding}
Peterson, J., Pearce, P.F., Ferguson, L.A., Langford, C.A.: Understanding
  scoping reviews: Definition, purpose, and process. Journal of the American
  Association of Nurse Practitioners  \textbf{29}(1),  12--16 (2017)

\bibitem{prasetya2020aplib}
Prasetya, I., Dastani, M., Prada, R., Vos, T.E., Dignum, F., Kifetew, F.:
  Aplib: Tactical agents for testing computer games. In: International Workshop
  on Engineering Multi-Agent Systems. pp. 21--41. Springer (2020)

\bibitem{procci2012measuring}
Procci, K., Singer, A.R., Levy, K.R., Bowers, C.: Measuring the flow experience
  of gamers: An evaluation of the dfs-2. Computers in Human Behavior
  \textbf{28}(6),  2306--2312 (2012)

\bibitem{rivero2017systematic}
Rivero, L., Conte, T.: A systematic mapping study on research contributions on
  ux evaluation technologies. In: Proceedings of the XVI Brazilian Symposium on
  Human Factors in Computing Systems. pp. 1--10 (2017)

\bibitem{saariluoma2014emotional}
Saariluoma, P., Jokinen, J.P.: Emotional dimensions of user experience: A user
  psychological analysis. International Journal of Human-Computer Interaction
  \textbf{30}(4),  303--320 (2014)

\bibitem{sanchez2012playability}
S{\'a}nchez, J.L.G., Vela, F.L.G., Simarro, F.M., Padilla-Zea, N.: Playability:
  analysing user experience in video games. Behaviour \& Information Technology
   \textbf{31}(10),  1033--1054 (2012)

\bibitem{stahlke2018usertesting}
Stahlke, S.N., Mirza-Babaei, P.: Usertesting without the user: Opportunities
  and challenges of an ai-driven approach in games user research. Computers in
  Entertainment (CIE)  \textbf{16}(2),  1--18 (2018)

\bibitem{steunebrink2007logic}
Steunebrink, B.R., Dastani, M., Meyer, J.J.C., et~al.: A logic of emotions for
  intelligent agents. In: Proceedings of the National Conference on Artificial
  Intelligence. vol.~22, p.~142. Menlo Park, CA; Cambridge, MA; London; AAAI
  Press; MIT Press; 1999 (2007)

\bibitem{steunebrink2008formal}
Steunebrink, B.R., Meyer, J.J.C., Dastani, M.: A formal model of emotions:
  Integrating qualitative and quantitative aspects. In: Dagstuhl Seminar
  Proceedings. Schloss Dagstuhl-Leibniz-Zentrum f{\"u}r Informatik (2008)

\bibitem{vermeeren2010user}
Vermeeren, A.P., Law, E.L.C., Roto, V., Obrist, M., Hoonhout, J.,
  V{\"a}{\"a}n{\"a}nen-Vainio-Mattila, K.: User experience evaluation methods:
  current state and development needs. In: Proceedings of the 6th Nordic
  conference on human-computer interaction: Extending boundaries. pp. 521--530
  (2010)

\bibitem{zhao2020winning}
Zhao, Y., Borovikov, I., de~Mesentier~Silva, F., Beirami, A., Rupert, J.,
  Somers, C., Harder, J., Kolen, J., Pinto, J., Pourabolghasem, R., et~al.:
  Winning is not everything: Enhancing game development with intelligent
  agents. IEEE Transactions on Games  \textbf{12}(2),  199--212 (2020)

\bibitem{zook2012temporal}
Zook, A., Riedl, M.: A temporal data-driven player model for dynamic difficulty
  adjustment. In: Proceedings of the AAAI Conference on Artificial Intelligence
  and Interactive Digital Entertainment. vol.~8 (2012)

\end{thebibliography}

\end{document}